\documentclass[12pt]{article}
\usepackage{epsfig}
\usepackage{amssymb}

\newcommand{\cZ}{{\cal Z}}
\newcommand{\Z}{{Z \!\!\! Z}}
\newcommand{\LL}{{I \!\! L}}
\newcommand{\beqn}{\begin{eqnarray}}
\newcommand{\eeqn}{\end{eqnarray}}
\newcommand{\eq}[1]{(\ref{#1})}
\newcommand{\const}{{\mathrm{const}}\,}
\newcommand{\dd}{\mathrm{d}}
\newcommand{\dD}{{\cal D}}
\newcommand{\dual}{\mbox{}^{\ast}}
\newcommand{\be}{\begin{equation}}
\newcommand{\ee}{\end{equation}}

\newcommand{\preprint}
{~\vspace{-1.5cm}
\begin{flushright}
{\large HU-EP-04/31},\hspace{5mm}
{\large KANAZAWA/2004-09}\\[1mm]
{\large LU-ITP 2004/017},\hspace{2mm}
{\large ITEP-LAT/2004-12}\\[3mm]
\end{flushright}}

\begin{document}
\baselineskip=14pt
\begin{center}

\preprint
{
{\Large\bf Monopole Chains in the Compact Abelian Higgs Model \\[2mm]
with doubly-charged Matter Field}

\vskip 1.0cm
{\bf \large M.~N.~Chernodub${}^{\bf \,a,b}$, R.~Feldmann${}^{\bf \,c}$,
E.-M.~Ilgenfritz${}^{\bf \,d}$, A.~Schiller${}^{\bf \,c}$}
\vskip 4mm
${}^{\bf a}\!$ {\it ITEP, B.Cheremushkinskaja 25, Moscow, 117218, Russia}\newline
${}^{\bf b}\!$ {\it Institute for Theoretical Physics, Kanazawa University, Kanazawa 920-1192, Japan} \newline
${}^{\bf c}\!$ {\it Institut f\"ur Theoretische Physik, Universit\"at Leipzig, D-04109 Leipzig, Germany} \newline
${}^{\bf d}\!$ {\it Institut f\"ur Physik, Humboldt-Universit\"at zu Berlin, Newtonstr. 15, D-12489 Berlin, Germany} \newline
}
\end{center}

\begin{abstract}
We study the properties of topological defects in the lattice compact
Abelian Higgs Model with charge $Q=2$ matter field. We find that
monopoles and antimonopoles form chain-like structures which are
dense in the confinement/symmetric phase. In this phase the mentioned
structures explain both the confinement of single-charged and the
breaking of strings spanned between doubly-charged test particles.
This observation helps to understand how the non-diagonal gluons,
once taken into consideration in the Abelian projection of gluodynamics,
could reproduce in this framework the string breaking for adjoint charges.
\end{abstract}

\section{Introduction}

In a series of papers, we have studied the three-dimensional compact
Abelian Higgs model (cAHM${}_3$) as a toy model~\cite{AHM} for
confinement and deconfinement in QCD. Starting from three-dimensional
compact QED (cQED${}_3$), where monopoles and antimonopoles in the
plasma phase are the agents of confinement at all
couplings~\cite{Polyakov}, we have studied deconfinement either
induced by rising temperature (in cQED${}_{2+1}$~\cite{cQED_3Tne0})
or occurring
due to interaction with a single-charged ($Q=1$) matter field in the
framework of the cAHM${}_3$. In the latter case, the presence of
matter fields eventually forces monopoles and antimonopoles to form
magnetically neutral bound states~\cite{AHM} connected by
Abrikosov-Nielsen-Olesen (ANO) vortices~\cite{Abrikosov}. This
mechanism of dipole formation is obviously different from that
observed in the high temperature phase of
cQED${}_{2+1}$~\cite{cQED_Binding,cQED_3Tne0}.

The role of monopoles is paralleled in four-dimensional Yang-Mills
theory where, in the dual superconductor scenario~\cite{dual_SC},
monopoles are also playing the role of agents of confinement (see,
{\it e.g.} the reviews~\cite{monopole_confinement}) and where the
presence of a dynamical matter field changes the monopole dynamics in
a way resulting in string breaking~\cite{string_breaking}. Not only
the similarities, also the differences to the four-dimensional
non-Abelian theory are interesting in this context. One has to
recognize that the dynamics of particle-like monopoles in four
dimensions is richer and changes at the deconfinement transition in a
way fundamentally different from the dynamics of instanton-like
monopoles in three dimensions.

In $3+1$ dimensional non-Abelian gauge theory,
the Abelian monopoles are condensed at low temperatures
and electric charges are confined~\cite{monopole_confinement}. However,
as the temperature rises the monopoles become massive and static and the
monopole condensate disappears. The static monopoles are unable to support
confinement of timelike moving charges.
However, such monopoles can still give rise to a non-zero spacelike string
tension.

In $2+1$ dimensional Abelian gauge theory,
the monopoles form a plasma phase
at low temperatures
in which the electric charges are confined similarly to $3+1$ dimensional theory.
Increasing temperature leads to the formation of
magnetic dipoles out of monopoles and antimonopoles. At sufficiently high
temperatures the monopole plasma is destroyed and a dilute dipole plasma is
formed. However, the dipole plasma is unable to create neither a ``spatial'' area
law nor the confinement of static charges~\cite{cQED_Binding,cQED_3Tne0}.

In four-dimensional non-Abelian theory there is another, rivalling
scenario of confinement based on center
vortices~\cite{old_vortex_picture}. A closer look reveals that it
does not invalidate the monopole picture of confinement but puts it
into another context, thus adding essential features to the monopole
dynamics. In short, percolation or static behavior of the latter
results from the respective properties of the vortices in the sense
of monopoles running inside the vortex sheets.
In the case of SU(2) this means that monopole worldlines are confined to
vortex sheets, and magnetic flux between opposite monopoles is
collimated inside the vortex sheet. This picture has been advocated
by Greensite {\it et al.}~\cite{Greensite_Zakopane} from the moment
that they revived the vortex scenario of confinement.
Later on, the picture helped to understand the different confinement properties
(and scales) of fundamental and adjoint charges which were difficult to explain
in the monopole picture~\cite{Greensite_Review}.

The cAHM${}_3$ is of interest to mimic this particular aspect due
to its non-perturbative features which arise from the presence of two
types of topological defects: monopoles and Abrikosov-Nielsen-Olesen
vortices~\cite{Abrikosov}. In the single-charged cAHM${}_3$ tight monopole
pairs are bound by ANO vortices. In three dimensions monopoles and
anti-monopoles
experience a logarithmically rising
attractive potential due to an anomalous dimension of
the photon propagator induced by the matter fields~\cite{Kleinert}.
The interaction guarantees the formation of monopole-antimonopole pairs.
When such pairs are forming, asymptotic confinement is simultaneously
lost for {\it all} test charges $q$. This is not the case we are now interested in.

In this paper we therefore consider the cAHM${}_3$ with
doubly-charged ($Q=2$) Higgs fields. Besides of its importance in
condensed matter physics~\cite{Q2_phase_structure} this model is
attractive in our context because it is clear~\cite{Fradkin_Shenker}
that the dynamical Higgs field can only screen external particles
with even charge $q$.
Before the transition to the Higgs phase sets in, the $q=1$ test charges
(the analogue of fundamental charges) should be confined at large separations, while
$q=2$ test charges (the analogue of adjoint charges) should suffer string breaking.
The assumed formation of {\it only} magnetic dipole states
is not able to explain this selective string breaking.
Finally, in the Higgs phase, none of the test charges will be
confined, which corresponds with the formation of monopole and
vortex clusters with a size of the order of the lattice spacing. Such
clusters may consist of pure vortex rings or also they may contain
monopoles.

\section{Some analytic considerations}

To enable analytical considerations and to simplify the numerical
simulations, we study the model in the London limit where the Higgs
field $\Phi$ at site $x$ is represented only by its phase
$\Phi_x = \exp{(i \varphi_x)}$.
The Wilson-type action of the model with
$Q$-charged Higgs field is
\beqn
  S_W[\theta,\varphi] = - \beta \sum_P \cos (\dd \theta)_P
- \kappa \sum_l \cos (\dd\phi - Q \theta)_l
\,,
  \label{eq:Wilson-action}
\eeqn
where $\theta_l$ is the link angle representing the compact
gauge field and
$(\dd theta)_P$
is the plaquette angle representing its
curl. $\beta$ denotes the gauge coupling and $\kappa$ is the hopping
parameter. Here and below we use the compact notations of
differential forms on the lattice (see the second of
Refs.~\cite{monopole_confinement} for a review).

The monopoles and the ANO-vortices appear due to compactness of the
phase angles $\theta_l$ and $\varphi_x$, respectively. For the sake
of our arguments it is instructive to perform the
Berezinsky-Kosterliz-Thouless~\cite{BKT} (BKT) transformation in
order to rewrite the partition function of the cAHM$_3$ in terms of
those defects. We use the Villain-type action (with couplings
$\tilde{\beta}$ and $\tilde{\kappa}$) instead
of~(\ref{eq:Wilson-action}):
\beqn
  \cZ = \!\!\int\limits^\pi_{-\pi} \!\! \dD \theta \!\!
  \int\limits^\pi_{-\pi}  \!\! \dD \varphi \!\!
  \sum_{n(c_2)}\sum_{l(c_1)}\! {\mathrm e}^{ - \tilde{\beta} {||\dd \theta + 2 \pi n||}^2
  - \tilde{\kappa} {||\dd \varphi -  Q\, \theta + 2 \pi l||}^2} ,
\label{eq:Z-fields}
\eeqn
where $n$ and $l$ are integer-valued forms
living on plaquettes $c_2$ and links $c_1$, respectively.

Applying the BKT transformation~\cite{BKT} (see also the second of
Refs.~\cite{monopole_confinement}) with respect to the gauge and the
Higgs fields and integrating them out we rewrite the partition
function~\eq{eq:Z-fields} as
\beqn
  \cZ \propto \cZ_{\mathrm{d}} = \!\!\sum\limits_{\dual m (\dual c_3)} \
  \sum\limits_{\stackrel{\dual j_m(\dual c_2)}{\delta \dual j_m = Q \, \dual m}}\
  {\mathrm e}^{- S_{\mathrm{d}}(\dual m, \dual j_m)}
  \label{eq:Z-objects}
\eeqn
with the defect action\footnote{In the journal version of this article the ${(\Delta + M^2)}^{-1}$
operator in the first term of the {\it r.h.s.} is inadvertently written as ${\Delta}^{-1}$.}
\beqn
  S_{\mathrm{d}}(\dual m, \dual j_m)
   = 4 \pi^2 \tilde{\beta} \, (\dual m, \frac{1}{\Delta + M^2} \dual m)
  + 4 \pi^2 \tilde{\kappa} \, (\dual j_m, \frac{1}{\Delta + M^2} \dual j_m) \; .
\label{eq:S-objects}
\eeqn
Here $M = Q
{(\tilde{\kappa}/\tilde{\beta})}^{1/2}$ is the tree-level mass of the
gauge boson, $\Delta$ is the lattice Laplacian. The integer-valued
forms $\dual m$ and $\dual j_m$ represent monopoles and vortices,
respectively, living on sites and links of the dual lattice. A site
$\dual c_3$ (link $\dual c_2$) of the dual lattice is dual to a cube
$c_3$ (plaquette $c_2$) of the original lattice.
The constraint in the second sum of Eq.~\eq{eq:Z-objects} requires
that the vortices begin (and end) at the monopole (anti-monopole) positions,
$\delta \dual j_m = Q \, \dual m$.
This indicates that a vortex carries a fraction $1/Q$ of the total magnetic
flux emanating from a monopole.

Let us now consider the contribution of vortices and monopoles to the
potential $V_{q}(R)$ between a pair of external test particles with
charges $\pm q$, separated by a distance $R$. The potential is given
in terms of the average of the Wilson loop $W_{q}(R,T) = \exp {\{i q
\,(J,\theta)\}}$ in the form $V_{q}(R)= - T^{-1} \log \langle
W_{q}(R,T) \rangle+\const$. The current $J$ runs around a contour of
rectangular shape, $R \times T$ with $R \ll T$. Using the same
transformations which led us from Eq.~\eq{eq:Z-fields} to
Eq.~\eq{eq:Z-objects}, the vacuum expectation value of the Wilson
loop will emerge factorized as $\langle W_{q} \rangle = {\langle
W_{q} \rangle}_{\mathrm{ph}} \cdot {\langle W_{q} \rangle}_{\mathrm{d}}$.
The first factor is simply the perturbative self-interaction of the
external loop \beqn
 {\langle W_{q} \rangle}_{\mathrm{ph}} \propto
\exp \Bigl\{ - \frac{q^2}{4 \tilde{\beta}} \, (J,\frac{1}{\Delta +
M^2} J) \Bigr\} \eeqn due to massive photon exchange. The
non-perturbative factor is due to topological defects: \beqn
  \langle W_{q} \rangle_{\mathrm{d}} =
  \cZ_{\mathrm{d}}^{-1} \!\!\!\!
  \sum\limits_{\dual m (\dual c_3)}
  \sum\limits_{\stackrel{\dual j_m(\dual c_2)}{\delta \dual j_m = Q \, \dual m}}\
  {\mathrm e}^{ - S_{\mathrm{d}}(\dual m,\dual j_m)
    - S_{\mathrm{int}}(\dual m,\dual j_m;J)}  ,
\label{eq:Wd} \eeqn with
\beqn
  S_{\mathrm{int}}(\dual m, \dual j_m; J) =
  - 2 \pi i \, \frac{q}{Q} \, (\dual \dd \dual  j_m, \frac{1}{\Delta + M^2} J)
  + 2 \pi i \, \frac{q}{Q} \, \LL(\dual j_0,J) \;.
\eeqn
The first term in the interaction $S_{\mathrm{int}}$ is a
Yukawa-type interaction (which reduces in the limit $M \to 0$ to a
usual cQED-like interaction) between a monopole and the charged test
particle. The second term is given by the linking number $\LL(\dual
j_0,J) = ( \dual \dd \dual j_0, \Delta^{-1} J) \in \Z$ between the
external particle trajectory $J$ and the closed part~\footnote{A
separation of a general vortex ensemble, $\dual j_m = \dual j_0 +
\dual j_m'$, into closed vortices, $\delta \dual j_0 = 0$, and open
ones, $\delta \dual j_m^{\prime} = Q \, \dual m$, is ambiguous.
However, in the sum~\eq{eq:Wd} the ambiguity disappears.} $\dual j_0$
of the vortex ensemble $\dual j_m$. The linking number describes the
Aharonov-Bohm (AB) topological interaction~\cite{MIP_Wiese_Zubkov}.

\begin{figure}[!htb]
  \begin{center}
     \includegraphics[scale=0.6,clip=true]{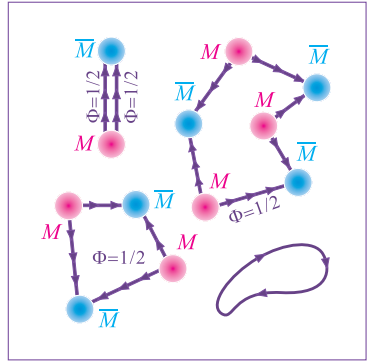}
   \end{center}
   \vskip -4mm
   \caption{A schematic view of simplest vortex-monopole configurations
   and a pure ANO vortex configuration in the cAHM${}_3$ with $Q=2$
   dynamical Higgs field.}
   \label{fig:monopoles:strings:Q2}
\end{figure}

The contribution of vortices to the potential $V_{q}$ is twofold. At
first, and most importantly, the closed vortices interact with the
electrically charged external particles via the AB effect if $q/Q
\notin \Z$. Secondly, the vortices influence the monopole dynamics
via the constraint, contributing in this way to the potential
indirectly, via the monopoles and their interaction with the test
particles. Since there are $Q$ vortices attached to each
(anti-)monopole, the vortices force the monopoles to form
magnetically neutral states which are dipoles for $Q=1$. If $Q=2$,
however, the monopoles may form not only dipoles but also extended
loop-like structures depicted in
Figure~\ref{fig:monopoles:strings:Q2}.
A typical loop structure contains vortex lines which, in general,
are non-orientable closed loops consisting of vortex segments with
definite orientations of magnetic fluxes.
A junction of
segments with different flux orientations must contain a monopole.
Thus non-orientable vortex loops should necessarily contain magnetic
monopoles.

Orientable vortices with magnetic flux cannot lead to confinement of
electric charges~\cite{Gubarev_AHM}. Indeed, the long-range
interaction between such vortices and the Wilson loop may be provided
only by the AB effect in a form visible in Eq.~\eq{eq:Wd}
(here we implicitly lean upon that we are working in a model with a mass
gap in which any interactions -- except for the topological ones like
via the linking number -- must be suppressed).
However, numerical simulations of the {\it non-compact} version of the
AHM${}_3$ show~\cite{Gubarev_AHM} that a confinement phase is absent
while the AB-interaction does exist. The vortices in this model can only
be orientable since monopoles are absent.

On the other hand, in the Maximal Center Gauge of the four-dimensional
Yang-Mills theory (YM${}_4$) the so-called center vortices are known
to lead to an area law of the fundamental Wilson loop via the linking
number representing the AB effect~\cite{Greensite_Zakopane,Greensite_2}.
The center vortices are non-orientable and therefore there is no contradiction
with the conclusions of Ref.~\cite{Gubarev_AHM}. Moreover, considered in an
Abelian sense
the non-orientable center vortices can be represented as segments of
orientable vortices with
monopoles separating the segments with different flux orientations
from each other, similarly to what we will find in the doubly-charged
cAHM${}_3$.~\footnote{Thus, one can ascribe the leading role in the
YM${}_4$ confinement phenomenon solely to the vortices
(as in Ref.~\cite{Greensite_Zakopane,Greensite_2}) but in this case the
vortices must be non-orientable which implies the existence of monopoles
living on the vortex sheets.}

We conclude that in the doubly-charged cAHM${}_3$ both monopoles and
vortices are responsible for the confinement of $q=1$ charges and the
breaking of the string between $q=2$ charges.
Note that in the single-charged cAHM${}_3$ non-orientable vortices
do not exist, the AB effect is absent and the vortices do not
feature in producing the confining forces.
In this case monopoles alone are responsible for confinement while
the vortices can only create
magnetic dipoles and, in this way, destroy confinement.

\section{Numerical simulations}

To check these general ideas, we have performed a numerical study of the
confining or nonconfining properties of electrically charged test particles
in the doubly-charged
cAHM${}_3$. The phase structure of the model was investigated in
Refs.~\cite{Fradkin_Shenker,Q2_phase_structure}. Before the Higgs phase sets
in, external particles with charge $q=1$ must be linearly confined in this
model whereas particles with charge $q=2$ show a flattening of the potential
at a certain distance. In order to observe this effect we have simulated the
model~\eq{eq:Wilson-action} on a $32^2\times 8$ lattice ($L_t=8$).
The choice of the asymmetric lattice is dictated by the fact that in the
case of symmetric lattices the potential can be measured only using
Wilson loops which are not suitable for an observation of the string
breaking effect. We have performed simulations at fixed gauge
coupling constant, $\beta=1.2$, the choice of which was motivated by
visualization reasons: to see clearly the monopole structures the
density of the monopoles must be neither too high nor too low.  We
have used about $10^4$ independent measurements for each value of the
hopping parameter~$\kappa$.

\begin{figure}[!htb]
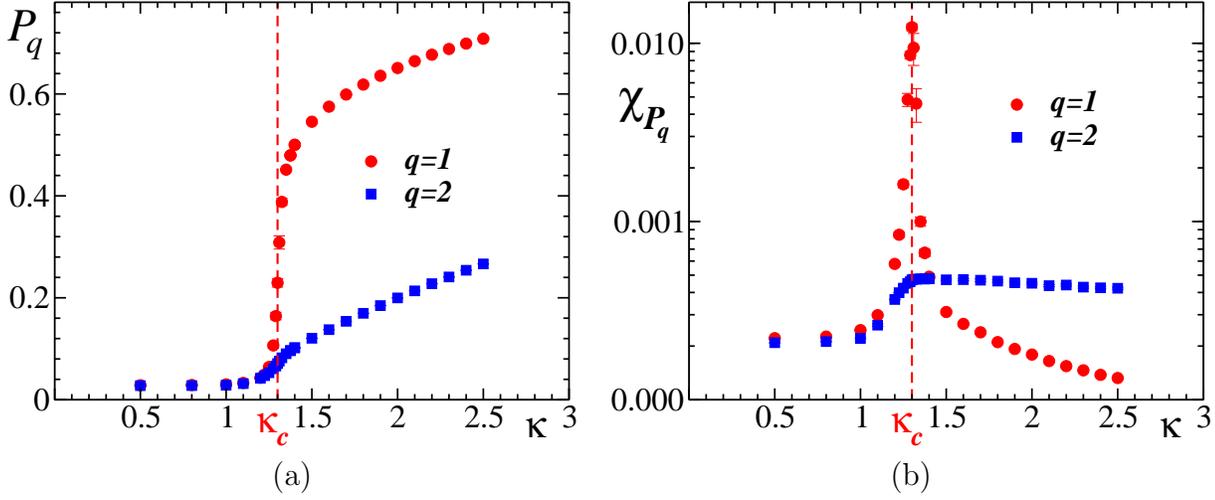

  \begin{center}
   \begin{tabular}{cc}
     \includegraphics[scale=0.35,clip=true]{polyakov.loop.eps} &
     \includegraphics[scale=0.35,clip=true]{polyakov.susc.eps} \\
     (a) & (b) \\
   \end{tabular}
  \end{center}
  \vskip -6mm
  \caption{(a) The $q=1,2$ Polyakov loops and (b) their susceptibilities
  $vs.$ $\kappa$.}
  \label{fig:polyakov:ls}
\end{figure}

Defining a Polyakov loop of charge $q$ at position $(x_1,x_2)$ via
\beqn
P_q(x_1,x_2)=\exp \Bigl\{i q \sum_{x_t=1}^{L_t}
\theta_3(x_1,x_2,x_t) \Bigr\}
\eeqn
with the obvious notation
$\theta_l= \theta_\mu (x_1,x_2,x_t)$ we show the {\it v.e.v.} of the
position average $P_q$ in Figure~\ref{fig:polyakov:ls}(a)
as a function of the hopping parameter $\kappa$. At small (large)
$\kappa$ the $v.e.v$'s of both loops are low (high) which corresponds
to the confinement (Higgs) phase. One can clearly see that the
average Polyakov loop for the $q=1$ external charges, which are
really confined on the left of the transition, is more sensitive to
the transition than $P_2$. The much less sensitivity of $P_2$ is
consistent with the observation of the $q=2$ string breaking (due
to $Q=2$ particles popping up in the vacuum) discussed below.

In Figure~\ref{fig:polyakov:ls}(b) we show the Polyakov loop
susceptibilities {\it vs.} $\kappa$. The susceptibility $\chi_{P_2}$
related to $q=2$ external particles changes between roughly constant
levels in the two phases at the transition point. At the same time,
$\chi_{P_1}$ shows a clear maximum signaling the transition from
confinement to deconfinement. We have fitted the susceptibility
$\chi_{P_1}$ in the vicinity of the maximum by the function
$\chi^{\mathrm{fit}}_{P_1} = C_1/{(C_2 +{(\kappa -
\kappa_c)}^2)}^\gamma$ with $C_{1,2}$, $\gamma$ and $\kappa_c$ being
fit parameters and we localize the transition at $\kappa_c =
1.300(1)$.

\begin{figure}[!htb]
  \begin{center}
    \includegraphics[scale=0.35,clip=true]{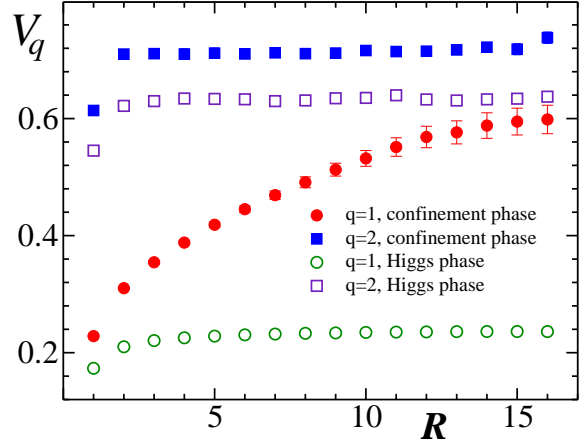}
  \end{center}
  \vskip -4mm
  \caption{
  The $q=1,2$ potentials in the confinement ($\kappa=1.275$)
  and the Higgs ($\kappa=1.325$) phases.}
  \label{fig:polyakov:correlators}
\end{figure}

The potentials
\beqn
V_{q}(|{\vec R}|) = - \frac{1}{L_t} \, \log \, \Bigl\langle
P_{q}({\vec 0}) P^{*}_{q}({\vec R}) \Bigr\rangle \; ,
\eeqn
extracted from the Polyakov loop correlators are presented in
Figure~\ref{fig:polyakov:correlators}
for both values of the electric test charges $q=1,2$. We show the
potentials both in the confinement (at $\kappa=1.275$) and the Higgs
phase (at $\kappa=1.325$). In the confinement phase the potential for
the $q=1$ external charges is
approximately linearly rising at large distances (the
flattening for $R \to 16$ is a result of lattice periodicity). The
$q=2$ potential shows a rapid flattening corresponding to the
dynamical $Q=2$ particle creation from the vacuum and, eventually, to
the string breaking already in the confinement phase. In the Higgs
phase {\it all} potentials show flattening due to deconfining nature
of this phase.

Let us now focus on the topological defects which should explain
this. The simplest characteristic of a topological defect is its
density. The monopole and the vortex densities are
\beqn
\rho_{\mathrm mon} = \frac{1}{V} \sum_{\dual c_3} |\dual m| \;,
\qquad \rho_{\mathrm vort} = \frac{1}{3 V} \sum_{\dual c_2} |\dual j_m| \;,
\eeqn
respectively.  The monopole charge is defined in the standard way,
\beqn
m = \frac{1}{2 \pi} \dd {[\dd \theta]}_{2\pi} \;,
\eeqn where
${[\cdots]}_{2\pi}$ denotes the integer part modulo $2 \pi$. The
vortex current is, following Ref.~\cite{ANO_definition}, defined as
\beqn
j = \frac{1}{2 \pi} (\dd {[\dd \varphi -
Q \theta]}_{2\pi} + Q {[\dd \theta]}_{2\pi}) \;.
\eeqn

\begin{figure}[!htb]
  \begin{center}
    \includegraphics[scale=0.35,clip=true]{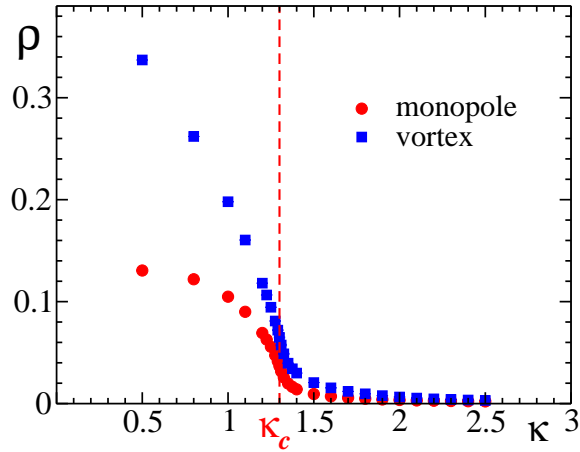}
  \end{center}
  \vskip -4mm
  \caption{The monopole and vortex densities $vs.$ $\kappa$.}
  \label{fig:monopole:vortices}
\end{figure}

According to Figure~\ref{fig:monopole:vortices}
the densities (shown in lattice units) of the monopoles and vortices
are gradually decreasing functions of $\kappa$ already in the
confinement phase. Towards the Higgs phase, the density of monopoles
and vortices drops faster, a fact which agrees with the expectation
that the confining properties of the cAHM${}_3$ are due to the
topological defects.

One has to realize that in the confinement phase the monopoles cannot
be in a plasma state because in this case the potentials for both
external charges $q=1,2$ would have to be linearly rising at large
separations, contrary to our observation
(Figure~\ref{fig:polyakov:correlators}). On the other hand, the
monopoles cannot form magnetically neutral monopole-antimonopole
bound states (dipoles) exclusively because in this case both $q=1,2$
potentials $V_q$ would have to show flattening -- essentially for
distances larger than the dipole size -- again in contradiction to
the observed behavior (Figure~\ref{fig:polyakov:correlators}). Thus
the only possible kind of monopole configurations -- which could
explain both the linearly rising potential for the $q=1$ electric
charges and the string breaking for the $q=2$ charges -- is a
monopole chain schematically plotted in
Figure~\ref{fig:monopoles:strings:Q2}.

\begin{figure}[!htb]
  \begin{center}
     \includegraphics[scale=0.75,clip=true]{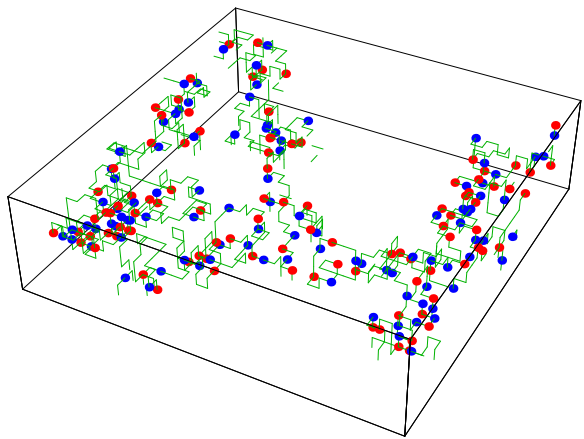}
  \end{center}
  \vskip -4mm
  \caption{Example of a monopole/vortex configuration in the
  confinement phase (at $\kappa=1.275$); monopoles are shown by circles,
  vortices by lines.}
  \label{fig:examples}
\end{figure}

If a vortex loop contains monopoles,
the monopoles and antimo\-no\-po\-les are mutually alternating along
the chain. Thus the magnetic flux coming from a monopole
inside the chain is separating into two parts, gradually squeezing
into vortices of finite thickness and, as a consequence, forming a
{\it non-orientable} closed magnetic flux. Each piece of such a
vortex carries in average a half flux, $\Phi = 2 \pi/Q \equiv \pi$.
If such a flux pierces the $q$-charged Wilson loop it provides a
contribution to the loop close to ${(-1)}^q$. This leads to the
necessary disorder for odd-charged external particles (eventually
leading to a linearly rising potential) whereas even-charged
particles are not confined. In Figure~\ref{fig:examples}
we visualize a typical monopole/vortex configuration observed in our
numerical simulations. This example shows the presence of the
monopole chains confirming the physical picture described above.

\begin{figure}[!htb]
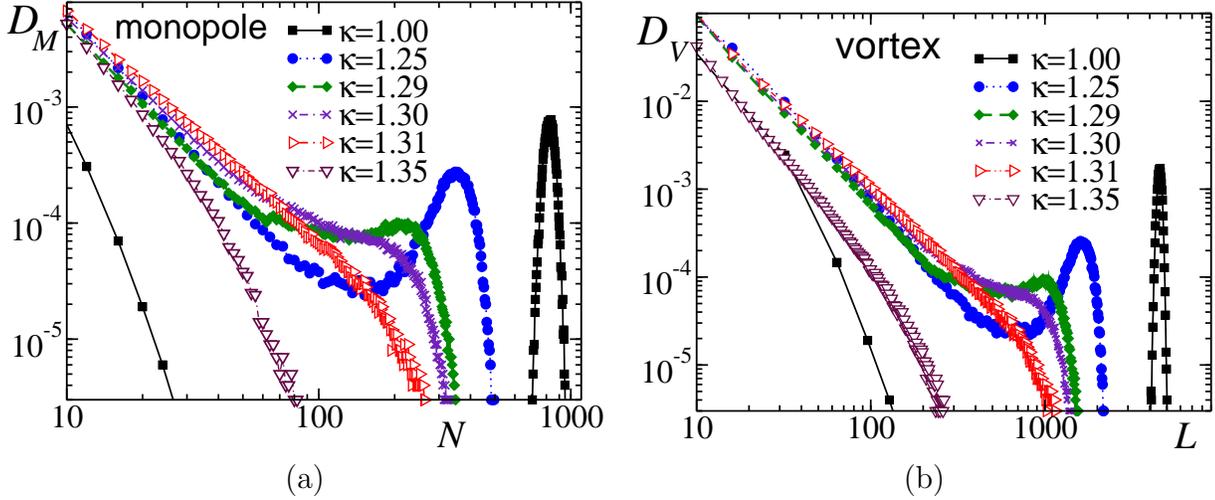

  \begin{center}
   \begin{tabular}{cc}
     \includegraphics[scale=0.35,clip=true]{kmonhist.eps} &
     \includegraphics[scale=0.35,clip=true]{kvorthist.eps} \\
     (a) & (b) \\
   \end{tabular}
  \end{center}
  \vskip -6mm
  \caption{Histograms of connected vortex clusters (as observed per configuration),
  (a) with respect to the monopole plus antimonopole number $N$ and
  (b) with respect to the vortex length $L$ (the number of links in the cluster),
  are shown for various values of $\kappa$. Confinement of $q=1$ test charges is
  associated with cluster percolation. }
  \label{fig:hist}
\end{figure}

One can suggest that in the confinement phase the monopole chains are
percolating as in Figure~\ref{fig:examples} (so that the monopoles
are relevant to infrared physics similarly to the monopoles in
QCD~\cite{monopole_confinement}) whereas in the Higgs phase the
monopole chains are relatively short. To check this idea we have
analyzed the cluster structure and show in
Figures~\ref{fig:hist}(a,b)
the normalized distributions of the observed mutually disconnected clusters
with respect to the number of monopoles and antimonopoles $N$ they
contain ($D_M$) and with respect to the number of vortices links (the
vortex length $L$) ($D_V$). One can see that in the confinement phase,
$\kappa < \kappa_c$, the distributions $D_M(N)$ and $D_V(L)$
have a two-peak structure. The first peak corresponds to the ultraviolet
clusters and is centered at small monopole numbers $N$
or small length $L$, respectively.~\footnote{Vortex loops without monopoles
obviously are not shown in the histogram of $D_M$.}
The second peak
is due to large (infrared) chain-like clusters which extend all over
the volume of the lattice. An example from the confining phase near
to the transition is shown in Figure~\ref{fig:examples}. In the
confinement phase the infrared peak in both distributions is clearly
present whereas in the Higgs phase the infrared peak is absent.
This is the signal of deconfinement.

\section{Discussion and outlook}

Summarizing, we have found that in the presence of a doubly-charged
Higgs field the monopoles must form chain-like structures. This
offers an explanation of both confinement of single-charged electric
particles and string breaking for doubly-charged test particles. This
physical picture -- observed in our study in the compact {\it Abelian}
gauge theory with a $Q=2$ dynamical Higgs field -- has a
close analogy with {\it non-Abelian} gluodynamics where tight
correlations between Abelian monopoles and center vortices (each in
the respective Abelian projection) have been
found~\cite{Greensite_3}.
Our observation also suggests a natural way for
the formation of monopole sheets (instead of chains) in gluodynamics.
For example, in the pure SU(2) gauge model (chosen here for
simplicity of discussion) the Abelian monopoles are defined with the
help of an Abelian gauge, in which the off-diagonal gluons
(originally ignored in the Abelian projection) play the role of the
{\it doubly-charged} matter fields coupled minimally to the leading
diagonal gluons. These matter fields may cause the
monopole trajectories to be confined inside sheets, a mechanism which
in turn should be responsible for the simultaneous occurrence (in the
pure gauge model) of confinement for fundamental charges (quarks) and
flattening of the potential between adjoint charges (gluons).
A more detailed study of monopole chain formation and confining properties
of the $Q=2$ model will be presented elsewhere~\cite{in_preparation}.

\section*{Acknowledgements}

We are grateful to J.~Greensite for useful comments to an earlier
version of the manuscript. This work is supported by grants RFBR 01-02-17456, DFG 436 RUS
113/73910, RFBR-DFG 03-02-04016, JSPS S04045 and MK-4019.2004.2.
E.-M.~I. is supported by DFG through the DFG-Forschergruppe ''Lattice
Hadron Phenomenology'' (FOR 465).

\end{document}